# An Analytical Framework for Analysis and Design of Networked Control Systems with Random Delays and Packet Losses

MishigaVallabhan[1], Seshadhri Srinivasan[2], S. Ashok[3], S. Ramaswamy*[4] and R. Ayyagari[5]

[1,3] Department of Electrical Engineering
National Institute of Technology- Calicut, Calicut, India

[2,4]Industrial Software Systems, ABB India Corporate Research Center, Bangalore, India-560068
srini@ieee.org

[5]Department of Instrumentation and Control Engineering
National Institute of Technology-Tiruchirappalli, Tiruchirappalli, India

***Abstract*—Delays and packet losses are undesirable from a control system perspective as they tend to adversely affect performance. Networked Control Systems (NCSs) are a class of control systems wherein control components exchange information using a shared communication channel. Delays and packet losses in the communication channels are usually random, thereby making the analysis and design of control loops more complex. The usual assumptions in classical control theory, such as delay free sensing and synchronous actuation, assume lesser significance when it comes to NCSs. Hence, this necessitates a reformulation / relook into the existing models used for NCS control loop analysis and design. In this paper, we study and present the reformulations required for NCSs to include random delays and packet losses in the channel. This paper therefore, provides a unified baseline and framework for analyzing a host of problems that can be captured as NCSs subjected to random delays and packet losses.**

## I. INTRODUCTION

Networked control systems (NCSs) contain a large number of interconnected devices that exchange data through shared communication channels. Recently, there has been an increasing interest among researchers in NCSs. A detailed review of NCSs alongside their applications can be found in ([1-11],[26-28], [30-31] and references therein). Two major challenges in analysis and design of NCSs are random delays and packet losses in the communication channel. Delays are undesirable as they not only degrade system performance but can also make an otherwise stable system, unstable. The effect of delays on NCSs was investigated in [13] and it has been shown that the NCSs performance and stability is affected by the delays in the communication channel. Packet losses result in system performance degradation and may result in loss of observability. Further, it is usually desirable from the control perspective to work with the most recent feedback information. This is not possible in the presence of packet losses. One may conclude from the above discussion that the usual assumptions from classical control theory, such as delay free sensing and synchronous actuation, are not entirely valid for NCSs. Hence, the first step to studying and analyzing NCSs with the intent to design controllers for them requires capturing the dynamics of the system alongside the communication constraints, vis-à-vis delays and packet losses.

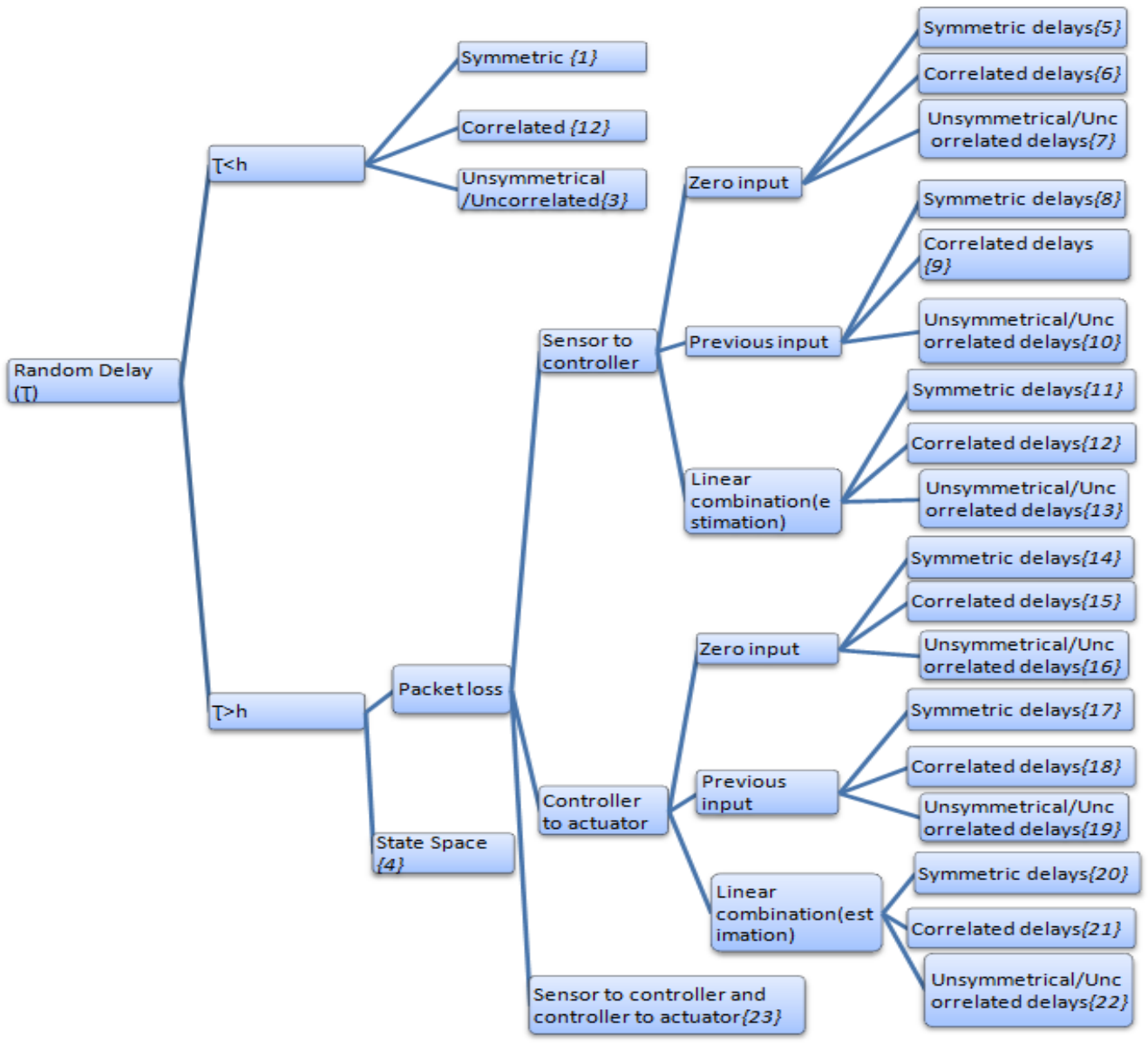

Figure 1: NCS Scenarios

Towards accomplishing this goal and given this scenario, a comprehensive framework for mathematical modeling of NCSs subjected to random delays and packet losses is paramount. In our work, we holistically consider these scenarios (as depicted in Figure 1) and propose a mathematical framework for analysis and design of NCSs.

A detailed review and research challenges alongside emerging applications for NCSs has been discussed in [8], wherein analysis and control of NCS subjected to random communication delays and packet losses has been identified as one of the potential future research areas. In [13], the authors investigated the effects of delays on NCSs and have also proposed controller design to compensate for the delays in the channel. Stochastic controller design for NCSs using (13) has been proposed in [14]. Peng et al. [18] designed state feedback control design for NCSs, the controller gains are computed using a Lyaponov analysis during each time epoch. In [6], modeling of NCSs subjected to random delays has been discussed. But, the analysis does not consider packet losses in the channel and relevant compensation methodologies. Further, the characteristics of the delays are not considered in the analysis.

Figure 2 **Research Timelines for different NCS models** shows the timeline of research efforts towards different mathematical models formulations for supporting real-time applications of NCSs.

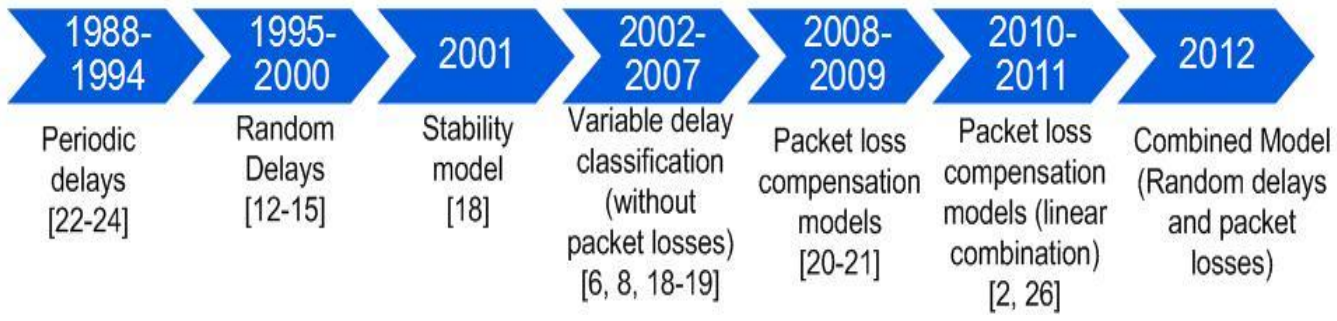

Figure 2 Research Timelines for different NCS models

Motivated by the above discussions, in this paper we propose to model the dynamics of NCSs subjected to random delays and packet losses. This helps in providing a comprehensive framework for mathematical modeling of NCSs. This work can be used as a baseline for future research in NCSs. The proposed mathematical model not only accounts for the random delays in the channel but also addresses various strategies to be employed for mitigating the effects of packet losses in the channel. We consider three widely adopted strategies, namely, *(i)* Transmitting zero in the event of a lost packet, *(ii)* Transmitting past value of control input in the event of dropped packet, and, *(iii)* Transmitting the estimate of the state or controller output. For the delays encountered in the channel, we consider broadly the two cases of delay: *(i)* the delay being less than the sampling time, and, *(ii)* delay being greater than the sampling time. Further, we analyze these classification based on the occurrence of the delay as: *(i)* delay in sensor to controller to be equal to the controller to actuator channel – termed synchronous channel delays, *(ii)* delay in sensor to controller channel being an integer or/and sub-multiple of the delay in the controller to actuator channel, and, *(iii)* delay in the controller to actuator channel being not correlated to the delay in the sensor to controller channel. This leads us to 23 different generic mathematical formulations for analysis and design of NCSs.

The rest of this paper is organized as follows. Section 2 presents the problem formulation and in section 3 a model of NCSs subjected to random communication delays is proposed. In section 4 we extend to the model of NCSs to account for the packet losses and different compensation schemes. Conclusions are drawn from the discussions in Section 5.

## II. Problem Formulation

Consider the system with dynamics

$$\dot{x}(t) = Ax(t) + Bu(t) \quad (1)$$

$$y(t) = Cx(t) \quad (2)$$

with a discrete state feedback controller

$$u(kh) = -Lx(kh) \quad (3)$$

Where $x \in R^n, u \in R^m, y \in R^p$ the state, input and output vectors respectively and $A \in R^{n\times n}, B \in R^{n\times m}, C \in R^{p\times n}$ are constant matrices of appropriate dimensions. Now consider sampling the continuous time system with a sample rate 'h', we have

$$x(kh+h) = \Phi(h)x(kh) + \Gamma(h)u(kh) \quad (4)$$

$$y(kh) = Cx(kh) \quad (5)$$

$$\Phi(h) = e^{Ah} \quad (6)$$

$$\Gamma(h) = \int_0^h e^{As}\, dsB \quad (7)$$

Now assume that there is a communication channel between sensor and controller *say* N1 .Let N2 be the communication channel between controller and actuator. The presence of communication channel induces delays $\tau_{sc}$ and $\tau_{ca}$ in the system as shown in Fig. 3 Generic block diagram of NCSs. The dynamics of the system (1) is then [12]:

$$x(kh+h) = \Phi(h)x(kh) + \Gamma_0(h,\tau)u(kh) + \Gamma_1(h,\tau)u(kh-h) \quad (8)$$

$$y(kh) = Cx(kh) \quad (9)$$

$$\Phi(h) = e^{Ah} \quad (10)$$

$$\Gamma_0(h,\tau) = \int_0^{h-\tau} e^{As}\, dsB \quad (11)$$

$$\Gamma_1(h,\tau) = e^{A(h-\tau)}\int_0^{\tau} e^{As}\, dsB \quad (12)$$

$$u(kh) = -L(h,\tau)x(kh) \quad (13)$$

Let us now assume that the packet loss in the system to be modeled as a binary random variable as in [29]:

$$\gamma_{sc} = \begin{Bmatrix} 0, packet\ loss \\ 1, no\ packet\ loss \end{Bmatrix} \quad (14)$$

$$\gamma_{ca} = \begin{Bmatrix} 0, packet\ loss \\ 1, no\ packet\ loss \end{Bmatrix} \quad (15)$$

Where $\gamma_{sc}$ and $\gamma_{ca}$ denotes the packet loss betwwen sentor to controller and controller to actuator respectively.

The dynamics of (3) with the packet loss included is then

$$x(kh+h) = \gamma_{sc}\big(\Phi(h)x(kh)\big) + \gamma_{ca}(\Gamma(h)u(kh)) \quad 16$$

Delays in NCSs are random and, as one may observe from (8), lead to time-varying system. Further, the packet losses need to be accounted in the system dynamics as in (16). This depends on the nature of delays and packet losses. The problem is to propose a mathematical framework for modeling NCSs subjected to random delays and packet loss.

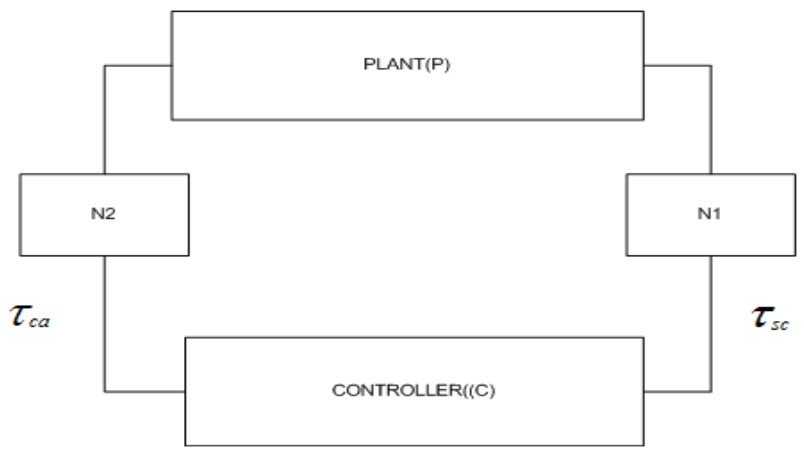

Fig. 3 Generic block diagram of NCSs

In our analysis, we propose mathematical framework for 23 different scenarios encountered in a NCSs.

## III. NCSs WITH RANDOM COMMUNICATION DELAYS

Communication delays in networks depend on the underlying protocol used. For eg, in CSMA/CD protocol (Network eg: Modbus over TCP/IP, Ethernet) delays are random in nature, in token passing and token ring protocol (Network eg:ControlNet), delays are bounded, CSMA/AMP protocol (Network eg , CAN, DeviceNet) has a constant delay in the network.

### A. *Modeling of NCSs with constant delay:*

There are various methods to handle constant delay like Pade's approximation, Nyquist analysis, or first order estimates. It may be seen that the mathematical model of NCSs subjected to constant delay is given by (8)-(13). Constant delays are common in NCSs integrated with token passing network like ControlNet [16].

### B. *Modeling of NCSs with time-varying delays:*

Time varying delays can be *(i)* less than sampling time and *(ii)* delays which is greater than sampling time. Varying delays are common in NCSs integrated with CSMA/CD networks like Ethernet, Modbus over TCP/IP [16]. In CSMA/CD based protocols transmitting nodes wait for random length of time (as determined by binary exponential backoff algorithm (BEB-algorithm)) for transmission access [16].

#### 1) *Delays less than sampling time (h)*

One may visualize that the total delay of the NCSs ($\tau_k$) at any time instant (k>0) is the sum of sensor to controller delay and controller to actuator delay.

$$\tau_k = \tau_{sc} + \tau_{ca}$$

We analyze the delays by considering the following three cases

1(a) $\tau_{sc} = \tau_{ca}$

When the delay between the sensor to actuator is equal, then the total delay $\tau_k$ is $2 * \tau_{ca}$

1(b) $\tau_{sc} = n * \tau_{ca}$

When the actuator to controller delay is an integral multiple or submultiples of sensor to controller delay.

1(c) $\tau_{sc} \neq \tau_{ca}$

When there is no correlation between the sensor to controller delay and controller to actuator delay. Mathematical model of NCSs subjected to random communication delay is given by

$$x(kh+h) = \Phi(h)x(kh) + \Gamma_0(h,\tau_k)u(kh) + \Gamma_1(h,\tau_k)u(kh-h) \quad (17)$$

$$y(kh) = Cx(kh) + Du(kh) \quad (18)$$

$$u(kh) = -L(h,\tau_k)x(kh) \quad (19)$$

$$\Phi(h) = e^{Ah} \quad (20)$$

$$\Gamma_0(h,\tau_k) = \int_0^{h-\tau_k} e^{As}\, dsB \quad (21)$$

$$\Gamma_1(h,\tau_k) = e^{A(h-\tau_k)} \int_0^{\tau_k} e^{As}\, dsB \quad (22)$$

#### 2) *Delays greater than sampling time (h)*

Network delays greater than sampling time *(h)* is considered as packet loss. If the time delay is longer than *h*,then the previous analysis has to be modified a little. If

$$\tau = (d-1)h + \tau' \quad (23)$$
$$0 < \tau' \le h$$

where *d* is an integer, the following equation is obtained:

$$x(kh+h) = \Phi x(kh) + \Gamma_0 u(kh-(d-1)h) + \Gamma_1 u(kh-dh) \quad (24)$$

$$\Phi(h) = e^{Ah} \quad (25)$$

$$\Gamma_0(h,\tau') = \int_0^{h-\tau'} e^{As}\, dsB \quad (26)$$

$$\Gamma_1(h,\tau') = e^{A(h-\tau')} \int_0^{\tau'} e^{As}\, dsB \quad (27)$$

The corresponding state-space description is

$$\begin{bmatrix} x(kh+h) \\ u(kh-(d-1)h \\ \vdots \\ u(kh-h) \\ u(kh) \end{bmatrix} = \begin{bmatrix} \Phi & \Gamma_1 & \Gamma_0 & \cdots & 0 \\ 0 & 0 & I & \dots & 0 \\ \vdots & \vdots & \vdots & \ddots & \vdots \\ 0 & 0 & 0 & \cdots & I \\ 0 & 0 & 0 & \cdots & 0 \end{bmatrix} \begin{bmatrix} x(kh) \\ u(kh-h) \\ \vdots \\ u(kh-2h) \\ u(kh-h) \end{bmatrix} + \begin{bmatrix} 0 \\ 0 \\ \vdots \\ 0 \\ I \end{bmatrix} u(kh) \quad (28)$$

$$y(kh) = Cx(kh) \quad (29)$$

$$u(kh) = -L(h,\tau)x(kh) \quad (30)$$

All three scenarios mentioned in 1(a), 1(b) and 1(c)is also valid for delays greater than the sampling period.

## IV. NCSs WITH RANDOM COMMUNICATION DELAYS AND PACKET LOSS

### A. *Random communication delay and packet loss between sensor and controller*

In order to compensate for the packet loss three strategies have been used in literature, they are: *(i)* transmitting zero, *(ii)* transmitting previous value of state and *(iii)* estimate of the state [20, 21]. Mathematical formulation considering packet loss compensation strategy is discussed in this section.

*1) Transmitting zero in the event of a lost packet*

Mathematical model for open loop is given as

$$x(kh+h) = \Phi(h)x(kh) \tag{31}$$

$$y(kh) = Cx(kh) \tag{32}$$

*2) Transmitting the previous state*

Mathematical model considering the transmission of the previous state is given as

$$\begin{aligned} x(kh+h) = \Phi(h)x(kh) + \gamma_{sc}(\Gamma_0(h,\tau_k)u(kh) \\ + \Gamma_1(h,\tau_k)u(kh-h)) \\ + (1-\gamma_{sc})\tilde{u}(kh) \end{aligned} \tag{33}$$

$$y(kh) = Cx(kh) \tag{34}$$

$$\Phi(h) = e^{Ah} \tag{35}$$

$$\Gamma_0(h,\tau_k) = \int_0^{h-\tau_k} e^{As}\, dsB \tag{36}$$

$$\Gamma_1(h,\tau_k) = e^{A(h-\tau_k)} \int_0^{\tau_k} e^{As}\, dsB \tag{37}$$

$$\tilde{u}(kh) = -L(h,\tau_k)x(kh-h) \tag{38}$$

$$\gamma_{sc} = \begin{cases} 0, packet\ loss \\ 1, no\ packet\ loss \end{cases}$$

*3) Transmitting the estimate of the state*

Mathematical model using estimated states when there is a packet loss between the controller and sensor is given as

$$\begin{aligned} x(kh+h) = \Phi(h)x(kh) + \gamma_{sc}(\Gamma_0(h,\tau_k)u(kh) \\ + \Gamma_1(h,\tau_k)u(kh-h)) \\ + (1-\gamma_{sc})\tilde{u}(kh) \end{aligned} \tag{39}$$

$$y(kh) = Cx(kh) \tag{40}$$

$$\Phi(h) = e^{Ah} \tag{41}$$

$$\Gamma_0(h,\tau_k) = \int_0^{h-\tau_k} e^{As}\, dsB \tag{42}$$

$$\Gamma_1(h,\tau_k) = e^{A(h-\tau_k)} \int_0^{\tau_k} e^{As}\, dsB \tag{43}$$

$$\begin{aligned} \tilde{x}(kh) = \alpha * \hat{x}(kh) + \beta * x(kh-h) \\ \hat{x}(kh) = \Phi(h)\hat{x}(kh-h) + \Gamma_0(h,\tau_k)u(kh-h) \\ + \Gamma_1(h,\tau_k)u(kh-2h) \\ \tilde{u}(kh) = -L(h,\tau_k)\tilde{x}(kh) \end{aligned} \tag{44}$$

Where $\alpha$ and $\beta$ are constants and its value varies between 0 and 1. The state is estimated using the linear combination [26] of present state estimate and the previous state.

The formulation for NCSs subjected to delays and packet loss as in (32) and (39) can be extended w.l.g. considering scenarios mentioned in 1(a), 1(b) and 1(c)in section 3. One may verify that the three scenarios can be captured using equations (32)-(44).

## B. *Random communication delay and packet loss between the controller and actuator*

In this section, we consider the packet loss in the communication channel between the controller and actuator. We consider three compensation schemes, they are: (i) transmitting zero, (ii) transmitting previous value of control input and (iii) estimate of the control input as in [20,21].

*1) Transmitting zero in the event of a lost packet*

The mathematical formulation of the scheme is given by equations (31)-(32)

*2) Transmitting the previous controller output*

Mathematical model considering the transmission of the previous control input is given as

$$\begin{aligned} x(kh+h) = \Phi(h)x(kh) + \gamma_{ca}(\Gamma_0(h,\tau_k)u(kh) \\ + \Gamma_1(h,\tau_k)\, u\,(kh-h)) \\ + (1-\gamma_{ca})u(kh-h) \end{aligned} \tag{45}$$

$$y(kh) = Cx(kh) \tag{46}$$

$$\Phi(h) = e^{Ah} \tag{47}$$

$$\Gamma_0(h,\tau_k) = \int_0^{h-\tau_k} e^{As}\, dsB \tag{48}$$

$$\Gamma_1(h,\tau_k) = e^{A(h-\tau_k)} \int_0^{\tau_k} e^{As}\, dsB \tag{49}$$

$$u(kh-h) = -L(h,\tau_k)x(kh-h)$$

$$\gamma_{ca} = \begin{cases} 0, packet\ loss \\ 1, no\ packet\ loss \end{cases}$$

*3) Transmitting the estimate of the control input*

Using estimated control input, we have

$$\begin{aligned} x(kh+h) = \Phi(h)x(kh) + \gamma_{ca}(\Gamma_0(h,\tau_k)u(kh) \\ + \Gamma_1(h,\tau_k)u(kh-h)) \\ + (1-\gamma_{ca})\hat{u}(kh) \end{aligned} \tag{50}$$

$$y(kh) = Cx(kh) \tag{51}$$

$$\Phi(h) = e^{Ah} \tag{52}$$

$$\Gamma_0(h,\tau_k) = \int_0^{h-\tau_k} e^{As}\, dsB \tag{53}$$

$$\Gamma_1(h,\tau_k) = e^{A(h-\tau_k)} \int_0^{\tau_k} e^{As}\, dsB \tag{54}$$

$$\begin{aligned} \hat{u}(kh) = \alpha * \tilde{u}(kh) + \beta * u(kh) \\ \tilde{u}(kh) = -L(h,\tau_k)\hat{x}(kh) \\ \hat{x}(kh) = \Phi(h)x(kh-h) + \Gamma_0(h,\tau_k)u(kh-h) + \\ \Gamma_1(h,\tau_k)u(kh-2h) \end{aligned} \tag{55}$$

Where $\alpha$ and $\beta$ are constants and its value varies between 0 and 1. The controller output is estimated using the linear combination of estimated control output and present controller output. The above formulation of NCSs considering packet loss in the communication channel between controller and actuator can also be extended w.l.g. to the three cases of delay discussed (1(a), 1(b) and 1(c)in section 3. Table I summarizes all the 23 scenarios considered for mathematical model. The schematic description of NCSs models discussed in this investigation are captured in Appendix I.

TABLE I
DIFFERENT SCENARIOS IN NCSs

| Case | Description |
|---|---|
| Case 1 | symmetric delays |
| Case2 | correlated delays |
| Case3 | unsymmetrical and uncorrelated delays |
| Case4 | state space model |
| Case5 | sensor to controller packet loss and symmetric delays with zero input strategy. |
| Case6 | sensor to controller packet loss and symmetric delays with previous input strategy |
| Case7 | sensor to controller packet loss and symmetric delays with linear combination strategy |
| Case8 | sensor to controller packet loss and correlated delays with zero input strategy |
| Case9 | sensor to controller packet loss and correlated delays with previous input strategy |
| Case10 | sensor to controller packet loss and correlated delays with linear combination strategy |
| Case11 | sensor to controller packet loss and unsymmetrical and uncorrelated delays with zero input strategy |
| Case12 | sensor to controller packet loss and unsymmetrical and uncorrelated delays with previous input strategy |
| Case13 | sensor to controller packet loss and unsymmetrical and uncorrelated delays with linear combination strategy |
| Case14 | controller to actuator packet loss and symmetric delays with zero input strategy |
| Case15 | controller to actuator packet loss and symmetric delays with previous input strategy |
| Case16 | controller to actuator packet loss and symmetric delays with linear combination strategy |
| Case17 | controller to actuator packet loss and correlated delays with zero input strategy |
| Case18 | controller to actuator packet loss and correlated delays with previous input strategy |
| Case19 | controller to actuator packet loss and correlated delays with linear combination strategy |
| Case20 | controller to actuator packet loss and unsymmetrical and uncorrelated delays with zero input strategy |
| Case21 | controller to actuator packet loss and unsymmetrical and uncorrelated delays with previous input strategy |
| Case22 | controller to actuator packet loss and unsymmetrical and uncorrelated delays with linear combination strategy |
| Case23 | sensor to controller packet loss and actuator to controller packet loss |

## V. CONCLUSION

Results derived in this paper can be used to model NCSs integrated with various communication protocols such as CSMA/CD, token ring, CSMA.AMP etc. One may verify that from equations (32)-(55), during computation of controller gains, delay in channel N2 is not available to the controller. It then becomes imperative to either estimate/determine the delay for computing the controller gains in order to design dynamic controllers. Most of the design methods available in literature devise a stabilizing controller for the worst case delay either using a Lyaponov formulation or a linear matrix inequality (LMI) based approach. It is seen that this methodology is conservative as the controller is always designed for worst case delay. Furthermore, the desired performance specifications may not always be met. The formulation provided above can be incorporated in controller design for NCSs subjected to random communication delays and packet losses for meeting desired performance specifications. An interesting extension of this study is to develop compensation methodologies for packet losses and dynamic controllers to meet performance specifications in the presence of random delays and packet losses. Further this paper can also serve as a handy reference to different the various scenarios that exists in NCSs.

## REFERENCE

[1] M. Bjorkbom, *Wireless control system simulation and network adaptive control*, PhD dissertation, School of Science and Technology, Department of Automation and Systems Technology, Altoo University, Oct. 2010.

[2] S. Seshadhri, *Control and estimation methodologies for networked control systems subjected to communication constraint*, PhD dissertation, Department of Instrumentation and Control Engineering, National Institute of Technology-Tiruchirappalli, India, Dec. 2010.

[3] J. P. Hespanaha, P. Naghshtabrizi, and Y. Xu, “A survey of recent results in networked control systems”, *Proceedings of IEEE*, vol. 95, no. 1, Jan. 2007, pp. 138-162.

[4] J. J.C. van Schendel, “Networked control systems: Simulation and Analysis”, *Traineeship report*, Technical University of Eindhoven, 2008.

[5] F-Y. Dang and D. Liu, “*Networked control systems: Theory and Applications*”, Verlag-Springer, 2008.

[6] P. Marti, J Yepez and J. M. Fuertes, “Control loop performance analysis over networked control system”, in *Proc . IEEEIECON’02*, vol. 4, Seville, Spain, Nov. 5-8, 2002, pp. 2880-2885.

[7] R. Olfati-Saber and R. M. Murray, “Consensus problems in networks of agents with switching topologies and time-delays”, *IEEE Transactions on Automatic Control*, vol.49, no. 9, 2004, pp. 1520-1533.

[8] J. Baillieul and P. J. Antsaklis, “Control and communication challenges in networked real time control systems”, in *Proc. IEEE*, vol. 95, no. 1, pp. 9-25, 2007.

[9] H. Chan and U. Ozguner, “Closed loop of control systems over communication systems with queues”, *Int. Journal of Control*, vol 62, no.3, pp. 493-510.

[10] Y. He, W. Min, L. Guo-ping , and S. Jin-Hua, “Output feedback stabilization for a discrete-time system with a time varying delay”, *IEEE Trans. Automatic Control*, vol. 53, no. 10, pp.2372-2377, 2008.

[11] S. Hu and Q. Ahu, ”Stochastic optimal control and analysis of stability of networked control systems with long delay”, *Automatica*, vol. 39, pp.1877-1884, 2003.

[12] K. J. Astrom and B. Wittenmark, *Computer controlled systems. Theory and Design* , 3rd ed. Upper Saddle River, NJ:Prentice-Hall, 1997.

[13] F. Lian, *Analysis , Design, Modelling and Control of Networked Control Systems* , PhD Dissertation, Department of Mechanical Engineering, University of Michigan, Ann Arbor , MI, USA, 2001.

[14] J. Nilsson, ”*Real Time Control Systems with Time Delays*”, PhD Dissertation, Department of Automatic Control, Lind Institute of Technology, Sweden, 1998.

[15] B. Wittenmark, J. Nilsson and M. Torngren, “Timing problems in the real-time control systems”, *Proc. American Control Conference*, vol. 3, pp. 2000-2004, 1995.

[16] F. Lian, J. R. Moyne and D. M. Tillbury, Performance evaluation of control networks :Ethernet, Control Net and DeviceNet, *IEEE Control System Magazine*, vol. 21, no. 1, pp. 66-83, 2001.

[17] Z. Wei, M. S. Branicky and S. M. Philips , “Stability of networked control systems :explicit analysis of delay”, *IEEE Control System Magazine*, vol. 21, no. 1, pp. 84-99, 2001.

[18] D. Yue, Q. Han and C. Peng, “State feedback controller design for network control system”, IEEE Trans.Circuits and Systems-II-*Express Briefs*, vol. 51, no. 11, pp. 640-644, 2004.

[19] Dimitrious Hristu- Varsakelis, W S Levine , ” Handbook of Networked Control and Embeded Control Systems”, Birkhauser, 2005.

[20] L. Schenato, "To zero or to hold control inputs in lossy networked control systems?", *European Control Conference(ECC07)*, 2007.

[21] L. Schenato, “Optimal estimate of networked control systems subjected to random delay and packet drop”, *IEEE Transaction on Automatic Control*, vol . 53, no. 5, June 2008.

[22] Y Haveli and A.Ray , “Intergrated communication and control system: Part 1 Analysis”, *Journal of dynamic system and measurement and control*, Dec-1988

[23] Y Haveli and A.Ray, “Intergrated communication and control system: Part I Analysis”, *Journal of dynamic system and measurement and control*, Dec-1988

[24] Y Haveli and A.Ray "Intergrated communication and control system: Part II Design", *Journal of dynamic system and measurement and control*, Dec-1988

[25] A. Ray, "Introduction to Networking for Integrated Control System", *IEEE Control Systems Magazine* , Jan 1989, pp-76-79.

[26] S. Seshadhiri, R Ayyangari, " Platooning over packet-dropping links", *International Journal of vehicle autonomous system* vol. 9, no. 12, pp. 46-62, 2011.

[27] X. Wan, H. Fang and S. Fu, "Fault detection for networked control system subjected to access constraints and packet loss", Journal of Systems Engineering and Electronics, vol. 22, no. 1, pp. 127-134, Feb 2011.

[28] Seshadhri, S., & Ayyagari, R. (2011). Dynamic controller for Network Control Systems with random communication delay. International Journal of Systems, Control and Communications, 3(2), 178-193.

[29] Srinivasan, S., & Ayyagari, R. (2010, October). Consensus algorithm for robotic agents over packet dropping links. In Biomedical Engineering and Informatics (BMEI), 2010 3rd International Conference on (Vol. 6, pp. 2636-2640). IEEE.

[30] Seshadhri, S., & Ayyagari, R. (2009, October). Hybrid Controllers for Systems with Random Communication Delays. In ARTCom (pp. 954-958).

[31] D. Peng, H. Zang, J. Lin, H. Li and F. Xia, " Simulation research for networked cascade control system based on truetime", Worl cd conference on Intelligent control and Automation(WCICA), pp. 485-488,Aug 2011.

[32] B. Sinopoli, L. Schenato, M. Franceschetti, K. Polla, M. I. Jordan and S. S. Sastry. "Kalman Filtering with intermittent observations", IEEE Transaction on Automatic Control, vol. 49, no. 9, pp. 1453-1464, Sep 2004.

[33] M.S. Branicky, S. M. Philips, W, Zhang, "Stability of networked control systems: explicit analysis of delays", *Proc. Americal Control Conference*, vol. 4, pp. 2352-2357, June 2000.

[34] G. P. Liu, S. C. Chai, J. X. Mu, and D. Rees, "Networked predictive control of systems with random delay in signal transmission channels", *International Journal of System Science*, vol. 39, Aug 2008.

APPENDIX I

TABLE II

MATHEMATICAL MODEL SUMMARY

| **Case** | **Sketch** | **Mathematical Model** | **Remarks** |
|---|---|---|---|
| No Delay (Case 0) | 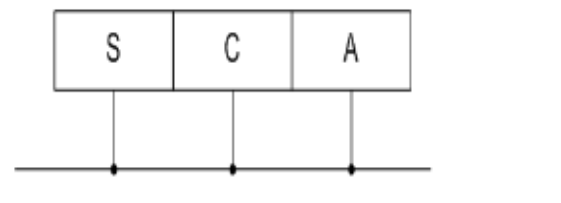 | (3)-(7) | |
| Constant Delay (Case 0) | 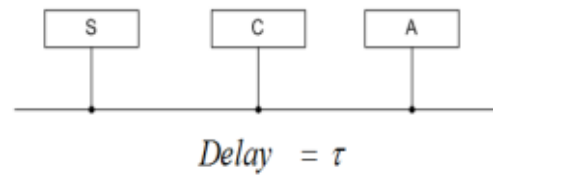 | (8)-(13) | |
| Case1 | 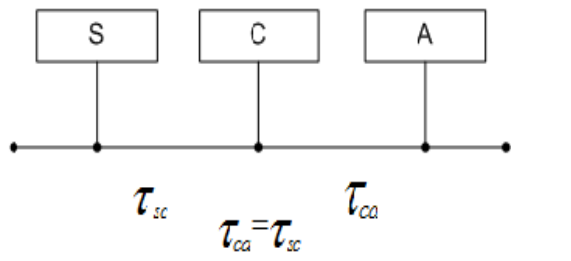 | (17)-(22) | |
| Case2 | 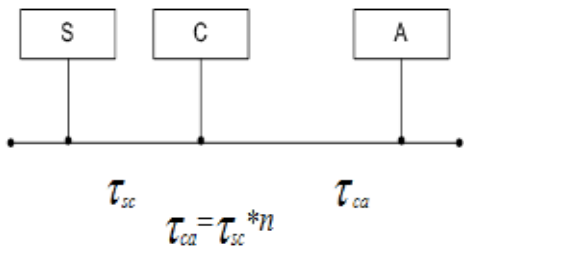 | (17)-(22) | |
| Case3 | 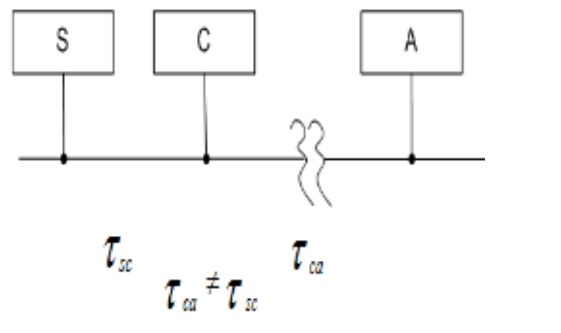 | | controller to actuator delay is not known(not possible to calculate $\tau_k$) |
| Case4 | | (28)-(30) | State space model |
| Case5-7 | 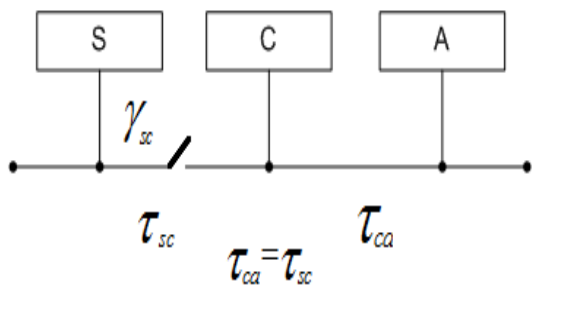 | (31)-(32) or (33-38) or (39-44) | 3 sets of mathematical model for packet loss compensation |
| Case8-10 | 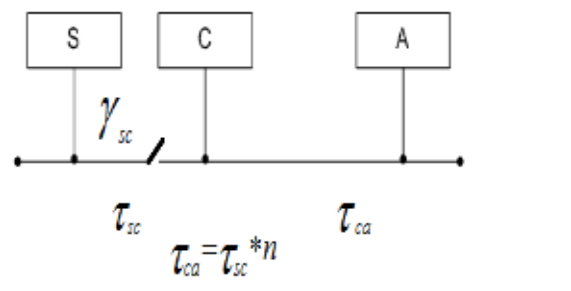 | (31)-(32) or (33-38) or (39-44) | 3 sets of mathematical model for packet loss compensation |
| Case11-13 | 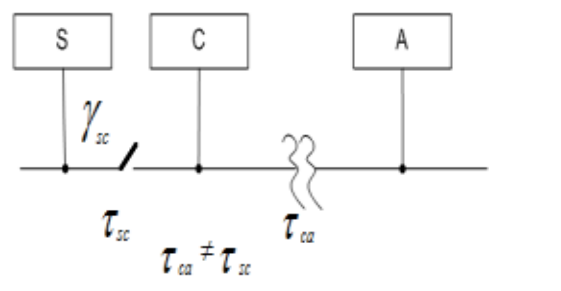 | -- | controller to actuator delay is not known(not possible to calculate $\tau_k$) |
| Case14-16 | 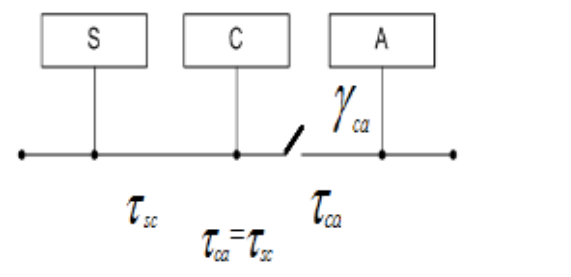 | (31)-(32) or (45)-(50) or (51)-(55) | 3 sets of mathematical model for packet loss compensation |
| Case17-19 | 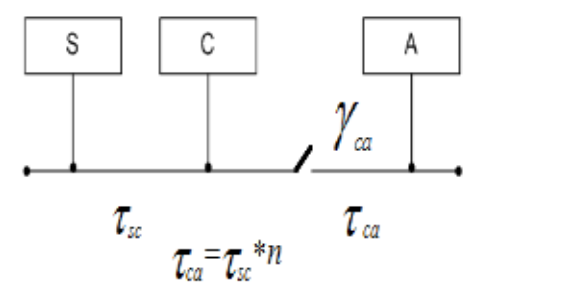 | (31)-(32) or (45)-(50) or (51)-(55) | 3 sets of mathematical model for packet loss compensation |
| Case20-22 | 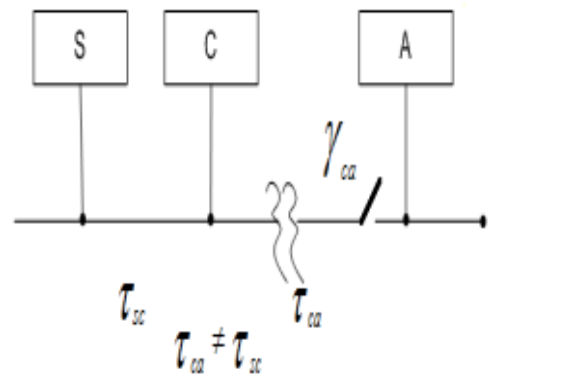 | | controller to actuator delay is not known(not possible to calculate $\tau_k$) |
| Case23 | 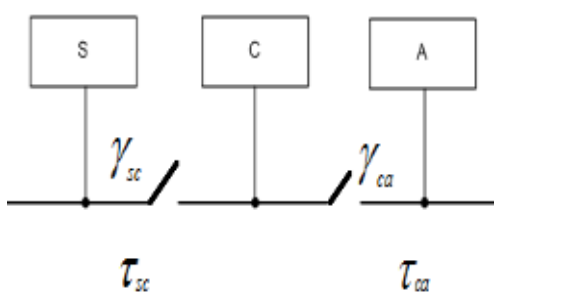 | | Non-deterministic |